\pdfoutput=1
\documentclass[acmsmall, nonacm]{acmart}
\usepackage{colortbl}

\usepackage[utf8]{inputenc}
\usepackage[T1]{fontenc}
\usepackage{microtype}

\usepackage{xargs}
\usepackage{lipsum}
\usepackage{xparse}
\usepackage{xifthen, xstring}
\usepackage[normalem]{ulem}
\usepackage{xspace}
\usepackage{float}
\makeatletter
\font\uwavefont=lasyb10 scaled 652

\newcommand\colorwave[1][blue]{\bgroup\markoverwith{\lower3\p@\hbox{\uwavefont\textcolor{#1}{\char58}}}\ULon}
\makeatother

\definecolor{pairedNegOneLightGray}{HTML}{cacaca}
\definecolor{pairedNegTwoDarkGray}{HTML}{827b7b}
\definecolor{pairedOneLightBlue}{HTML}{a6cee3}
\definecolor{pairedTwoDarkBlue}{HTML}{1f78b4}
\definecolor{pairedThreeLightGreen}{HTML}{b2df8a}
\definecolor{pairedFourDarkGreen}{HTML}{33a02c}
\definecolor{pairedFiveLightRed}{HTML}{fb9a99}
\definecolor{pairedSixDarkRed}{HTML}{e31a1c}

\graphicspath{{./images/}}

\makeatletter
\newcommand\requiredelimiter[2][########]{%
  \ifdefined#2%
    \def\@temp{\def#2#1}%
    \expandafter\@temp\expandafter{#2}%
  \else
    \@latex@error{\noexpand#2undefined}\@ehc
  \fi
}
\@onlypreamble\requiredelimiter
\makeatother

\newcommand\newdelimitedcommand[2]{
\expandafter\newcommand\csname #1\endcsname{#2}
\expandafter\requiredelimiter
\csname #1 \endcsname
}

\newdelimitedcommand{toolname}{Tool}

\usepackage{tikz}
\usetikzlibrary{arrows}
\usetikzlibrary{shapes}
\newcommand*\circled[1]{\tikz[baseline=(char.base)]{
            \node[shape=circle,fill=pairedOneLightBlue,inner sep=1pt] (char) {#1};}}

\usepackage{booktabs}   
\usepackage{subcaption} 

\usepackage{listings}
\setcopyright{none}
\acmConference[]{}{}{}
\acmYear{}
\acmISBN{}
\acmDOI{}
\startPage{1}

\begin{document}

\title[QSSA]{QSSA: An SSA-based IR for Quantum Computing}
\subtitle{}                             

\author{Anurudh Peduri}
\affiliation{
  \department{CSTAR}              
  \institution{IIIT Hyderabad}            
  \country{India}
}
\email{anurudh.peduri@research.iiit.ac.in}          

\author{Siddharth Bhat}
\affiliation{
  \department{CSTAR}              
  \institution{IIIT Hyderabad}            
  \country{India}
}
\email{siddharth.bhat@research.iiit.ac.in}          

\begin{abstract}


Quantum computing hardware has progressed rapidly. Simultaneously, there has
been a proliferation of programming languages and program optimization
tools for quantum computing.
Existing quantum compilers use intermediate representations (IRs) where quantum
programs are described as \emph{circuits}. Such IRs fail to leverage
existing work on compiler optimizations.
In such IRs, it is non-trivial to statically check for physical constraints
such as the no-cloning theorem, which states that qubits cannot be copied.
We introduce QSSA, a novel quantum IR based on static single assignment (SSA)
that enables decades of research in compiler optimizations to be applied to
quantum compilation.
QSSA models quantum operations as being side-effect-free. The inputs and
outputs of the operation are in one-to-one correspondence; qubits cannot be
created or destroyed. As a result, our IR supports a static analysis pass
that verifies no-cloning at compile-time. The quantum circuit is fully
encoded within the def-use chain of the IR, allowing us to leverage
existing optimization passes on SSA representations such as redundancy
elimination and dead-code elimination. Running our QSSA-based compiler on
the QASMBench and IBM Quantum Challenge datasets, we show that our optimizations perform
comparably to IBM's Qiskit quantum compiler infrastructure.
QSSA allows us to represent, analyze, and transform quantum programs using the
robust theory of SSA representations, bringing quantum compilation into the
realm of well-understood theory and practice.
\end{abstract}


\renewcommand\footnotetextcopyrightpermission[1]{} 

\maketitle

\section{Introduction}

Quantum computers harness quantum mechanics to perform efficient computation.
They are widely believed to be capable of ``quantum supremacy'' --- the
ability to perform computations that are not efficiently simulatable using
classical computation \cite{arute2019quantum}. Due to the rise of Noisy
Intermediate-Scale Quantum (NISQ) technology, where we possess quantum
computers that are sufficiently powerful to exhibit quantum supremacy
\cite{preskill2018nisq}, there is a growing interest in quantum compiler
toolchains that analyze and optimize quantum programs. Quantum computation
has quite different semantics from classical computation. Quantum programs
operate on quantum bits (qubits), the counterpart to a classical bit. The
state of a qubit can be in a superposition between the classical bit states
of $|0\rangle$, $|1\rangle$. Other differences from classical computing
include the fact that all quantum operations are \emph{reversible} and that
it is impossible to create an independent and identical copy of an arbitrary
unknown qubit, as given by the no-cloning theorem \cite{wootters1982single}.
These differences make it challenging to analyze and optimize quantum
programs.

Designing IRs for quantum programming languages is made more complex
due to \emph{hybrid quantum-classical programs}, where a
quantum circuit and a classical computer communicate information to perform
computations. To express such computations, we need a \emph{single} language to
express both quantum and classical computation. This is a challenging language
design problem that needs an IR design that allows quantum computing and
classical computing semantics to interoperate harmoniously in the same IR.
Two major classes of quantum programming languages exist. First,
quantum compilers for pure quantum programs, which use
intermediate representations (IR) where quantum programs are described as
\emph{circuits}. This choice is inspired by the quantum computing literature,
where quantum algorithms are commonly analyzed as circuits. The circuit
encoding poses two problems. (a) the IR exists as an in-memory DAG with no
textual representation. This makes testing, debugging, and interfacing with
these compilers challenging.  (b) this representation of quantum programs
as circuits cannot express hybrid computations. IRs such as
Qiskit~\cite{aleksandrowicz2019qiskit},
Cirq~\cite{cirq_developers_2021_4586899}, and XACC~\cite{xacc_2020}
take this choice to represent only pure quantum programs. To cover hybrid
quantum-classical programs, IRs such as QIR~\cite{QIR},
Scaffold~\cite{scaffcc}, QCL~\cite{omer_qcl}, and Quil~\cite{smith2016pyquil}
attempt to bridge the gap by representing the classical portion with a
standard SSA-based imperative IR, and the quantum portion as a special type
of ``qubit memory''. Instructions that modify qubits are represented as loads
and stores into the special qubit memory. The presence of memory operations
makes the IR impure due to side-effects. It disables the reuse of standard
optimizations across the classical and quantum parts of the IR due to the
presence of two different flavors of memory. Furthermore, the use of
memory-based qubit semantics forces the use of complicated alias and dataflow
analysis to reason about the IR. In summary, pure quantum circuit IRs lack
expressivity, while hybrid quantum circuit IRs lack optimisability.

For an optimizable hybrid quantum-classical program IR, we identify four key
requirements:
(1) The IR must represent hybrid quantum-classical
programs with complete encodings of both classical computation
and quantum computation. 
(2) The IR must naturally express well-known transformations. For
example, redundancy elimination is commonly run on the IR
for canonicalization as it is an \emph{enabling transformation}, which makes the IR
simpler. For such a transformation to depend on complex static analysis
makes the transformation both more expensive (since the static analysis must be
computed) and less robust (since the static analysis is an over-approximation).
(3) The IR should be as similar as possible to existing compiler IRs to (a)
enable the reuse of algorithms and code, and (b) to identify what extensions
are necessary for classical compiler IRs to express quantum programs naturally. 
Finally, to aid compiler debugging, development, and testing, (4) The IR must
be easily readable and modifiable for humans. This implies that the IR cannot
solely exist as in-memory data structures. The compiler toolchain must be able
to dump the in-memory IR into a readable textual format and ingest the
textual representation back into the compiler pipeline. This ensures that the
textual representation is lossless, which enables modular testing of the
compiler. This has been used to great success by LLVM's \texttt{FileCheck}
infrastructure to unit test the entire compiler.

To fulfill the above four criteria, We craft QSSA - an SSA-based IR - to
represent and optimize hybrid classical-quantum programs. QSSA expresses the
circuit's data dependencies with the def-use chain of SSA values.
Furthermore, since SSA names the outputs of each primitive instruction, it
effectively names every edge in the circuit DAG, which aids circuit
debuggability. For the hybrid quantum-classical regime, QSSA provides
side-effect free, value-semantics based operations for qubit manipulation.
This ensures that QSSA's qubit representation can interoperate with classical
SSA analyses. QSSA thus establishes the extensions for SSA necessary to
express hybrid programs. To ensure that quantum invariants are preserved, we
develop a static analysis that verifies that each qubit is used at most
once. QSSA extends the representation used between pure quantum
representations - a DAG-based, SSA-like encoding - to the full hybrid
quantum representation. We argue that to harness SSA and optimize hybrid
quantum-classical programs, we must extend SSA with qubits to
preserve value semantics. To our knowledge, this paper is the first to propose
this solution. We make the following contributions:
\begin{itemize}
    \item We define the semantics of quantum-SSA (QSSA), where we identify the
        extensions necessary for side-effect free SSA to naturally describe hybrid
        quantum-classical programs. 
    \item We present a novel algorithm to statically verify no-cloning for QSSA,
        which takes quadratic time on general control flow graphs, and linear time
        on control flow graphs that preserve high level program structure using
        \emph{regions}.
    \item We translate classical SSA analyses and transformations into the QSSA
        framework while maintaining quantum invariants.
    \item We develop a prototype implementation of our proposal in MLIR, a compiler
        development framework for SSA-based IRs that has builtin support
        for well-known SSA optimizations, along with an API design that
        ensures that QSSA has a textual on-disk representation which can be fed
        back into our compiler pipeline.
    \item We validate our implementation on the QASMBench \cite{li2020qasmbench}
        and IBM Quantum Challenge~\cite{ibmqxchallenge} datasets, and show that our
        optimizations perform comparably to IBM's Qiskit quantum compiler
        infrastructure.
\end{itemize}

\section{Background}

This section provides a broad overview of the ideas necessary to appreciate
QSSA. It introduces the SSA intermediate representation, the ideas of quantum
computing that influence the design of our SSA language, and the MLIR
compiler framework for building SSA-based compilers.

\subsection{Quantum Computing}
In this paper, we deal with discrete quantum systems, which are comprised of
\emph{qubits}. Quantum programs are built by applying \emph{gates} on these
qubits~\cite[Chapter~4]{chuangQIQC}.

\subsubsection{Qubits}
A Qubit is a two-level quantum system whose state is described by a unit
vector in a complex 2D Hilbert space $\mathbb{H}$. We consider $\{|0\rangle,
|1\rangle\}$ to be the canonical basis of this space. The general state of a
qubit is of the form $a |0\rangle + b |1\rangle$ where $a, b \in \mathbb C$,
$|a|^2 + |b|^2 = 1$. The basis states are analogous to classical bits, but
qubits are allowed to take on more states which are superpositions of these
basis states. The tensor product of individual qubit states represents the
state of a multi-qubit system.

\subsubsection{Gates} 
A gate is a unitary operator acting on the space of states. A gate is
analogous to a logical gate in classical computing. On applying a physical
gate on a qubit, its state vector gets transformed by the unitary operator.
The unitary condition ensures that the resulting state is also a unit vector.
Quantum gates can take in any number of qubits as input and return the same
number of qubits as output after applying the unitary. 

\subsubsection{Measurement}
To extract the state of a qubit, a \emph{measurement} is performed. The
result of measuring a qubit in the state $a|0\rangle + b|1\rangle$ is a
single bit ($0$ or $1$). This operation is \emph{probabilistic} and produces
the bit $0$ with probability $|a|^2$, and produces the bit $1$ with
probability $|b|^2$. The aforementioned condition that $|a|^2+|b|^2 = 1$ is a condition
that ensures that the total probability is always $1$. On measurement, the
qubit collapses to the measured state, and repeated measurements will give the
same result.

\subsubsection{The No Cloning theorem}
In physics, the no-cloning theorem \cite{wootters1982single} states that it is
impossible to create an independent and identical copy of an arbitrary unknown
quantum state. This implies that in the compiler IR, we must not allow programs
that require multiple copies of the same value. Existing compilers generally
enforce this by a run-time check.

\subsubsection{Hybrid Classical-Quantum Programs}
Quantum circuits consist of a sequence of gates applied on a set of qubits,
which are finally measured to extract some result. Frameworks like Cirq are
used to express pure quantum circuits. This design is restrictive, as it
does not allow any classical feedback to control which quantum gates to
apply.
Hybrid programs are allow the interleaving of classical computation control flow
with quantum operations. This helps describe a larger class of quantum
programs. One instance of this is the class of  \emph{Variational Quantum Algorithms}~\cite{cerezo2020variational},
which uses results of quantum measurements to perform classical computation,
which in turn decides which quantum operations to apply next, and so on.

\subsection{Quantum Intermediate Representations}
We compare our design against two standard quantum IRs: OpenQASM and QIR.
These represent extrema in the spectrum of quantum compilation. OpenQASM is
a minimal IR designed to represent fixed size, pure quantum circuits. QIR is the
IR for Q\#, a general-purpose hybrid quantum programming language.

\subsubsection{OpenQASM}~\footnote{In this paper, OpenQASM refers to version
2.0. At the time of writing, version 3.0 is under development, which does not 
change the core characteristics of the IR}
is a low-level quantum assembly language by
IBM~\cite{cross2017open}. It represents universal physical circuits over
the $CNOT$ plus $SU(2)$ basis with straight-line code. It has support for
measurement, reset, and gate subroutines. It uses memory semantics to
represent qubit manipulation. OpenQASM has been designed for the representation
and optimization of pure quantum circuits. Thus, OpenQASM acts as an anchor
throughout the paper to discuss pure quantum circuits and qubit-memory based
semantics.

We briefly describe its instructions that are used in the paper. Fixed-size
qubit registers are allocated using the \texttt{qreg} instruction. For
example, \texttt{qreg q[4];} declares a register of 4 qubits. Similarly,
\texttt{creg} allocates a fixed-size classical register of bits. A gate
application is like a function call - the name of the gate followed by qubit
arguments. For example, to apply the $CNOT$ gate, we write \texttt{cx q[0],
q[1];}. The \texttt{measure} instruction measures a qubit/qubit register and
stores it in a classical register. \texttt{measure q[0] -> c[0];} stores the
measurement result of \texttt{q[0]} in \texttt{c[0]}.

\subsubsection{QIR}
is an LLVM-based IR for hybrid quantum programs~\cite{QIR}. QIR is the compiler
intermediate target for Q\#, A hybrid quantum programming language by Microsoft.
QIR extends LLVM with opaque intrinsics to represent qubits and qubit arrays as opaque pointer types. 
QIR's encoding of LLVM does not interoperate with the rest of the LLVM 
language, as we will demonstrate in \autoref{subsec:redundancy-elimination}.
Thus, QIR acts as an anchor throughout the paper  to discuss hybrid
classical-quantum programs, as well as issues with mixing classical and qubit
memory semantics 

\subsection{Static Single Assignment}
An intermediate representation (IR) is in static single assignment form (SSA)
\cite{rastello2010ssa} if each variable is assigned exactly once and no
variable remains undefined. SSA has gained popularity in imperative compilers
such as GCC~\cite{stallman2002gnu}, LLVM~\cite{lattner2004llvm}, and many
others, as data-flow information is explicitly expressed through dependencies
from the definition of a value to its uses (def-use chains). SSA-based IRs
typically use basic blocks that hold lists of sequentially executed
operations, each taking a list of argument values and returning a tuple of
return values. Terminator operations at the end of each basic block, which
either branch to another basic block or return from a function, combine these
basic blocks into a control flow graph (CFG). While IRs typically use a flat
CFG, MLIR~\cite{lattner2021mlir} recently introduced nested control flow as a
first-class concept to support abstractions that require control over scope.
Operations can now receive \emph{regions}, which are nested single-entry
sub-CFGs, as additional arguments. Regions make it easy to express concepts
such as loop bodies or branches of an if-statement.

\subsection{MLIR}
MLIR is a new compiler infrastructure under the LLVM umbrella~\cite{lattner2021mlir}. It
aims at simplifying development domain-specific compilers. For this, MLIR
provides a non-opinionated, SSA-based intermediate representation (IR).
Having a customizable IR allows compiler developers to model domain-specific
concerns by introducing custom types, operations, and attributes. Here, we
briefly describe the relevant aspects of MLIR that are used in our compiler.

\subsubsection{Modules}
Each program in MLIR is called a \emph{Module}. A Module consists of several
global functions. Function names such as \texttt{@foo} are global and allow
for linking function calls across modules. A function consists of a sequence
of SSA operations. Each SSA value is local to the scope of the function and has
a name of the form \texttt{\%bar}.
\subsubsection{Operations}
SSA values are produced by \emph{Operations}, such as \texttt{addi}, which
stands for the integer addition operation. Each operation takes zero or more
SSA operands that are defined before it and returns zero or more SSA values.
Operations can also have compile time constants attached to them, such as
\texttt{\{phase = 90.0 : f64\}}. These are called \emph{attributes}.
\subsubsection{Dialects}
A \emph{Dialect} in MLIR is a collection of operations and types of a certain
kind. In this paper, we use two existing dialects, \texttt{std} and \texttt{scf}.
The \texttt{std} dialect contains all basic operations such as constant
declarations, arithmetic operations and memory manipulation. The Structured
Control Flow Dialect \texttt{scf} contains \texttt{if-else} and \texttt{for}
loop constructs.

\section{The QSSA Representation}

This section describes the QSSA Intermediate Representation, prototyped as an
MLIR dialect. The aims of the encoding are twofold: (1) Encode all 
qubit operations using value semantics, thereby making the program side
effect free, and (2) Exploit the SSA def-use chain to encode
the quantum circuit within the IR implicitly. 

\begin{figure}[!htb]
\includegraphics[width=\textwidth]{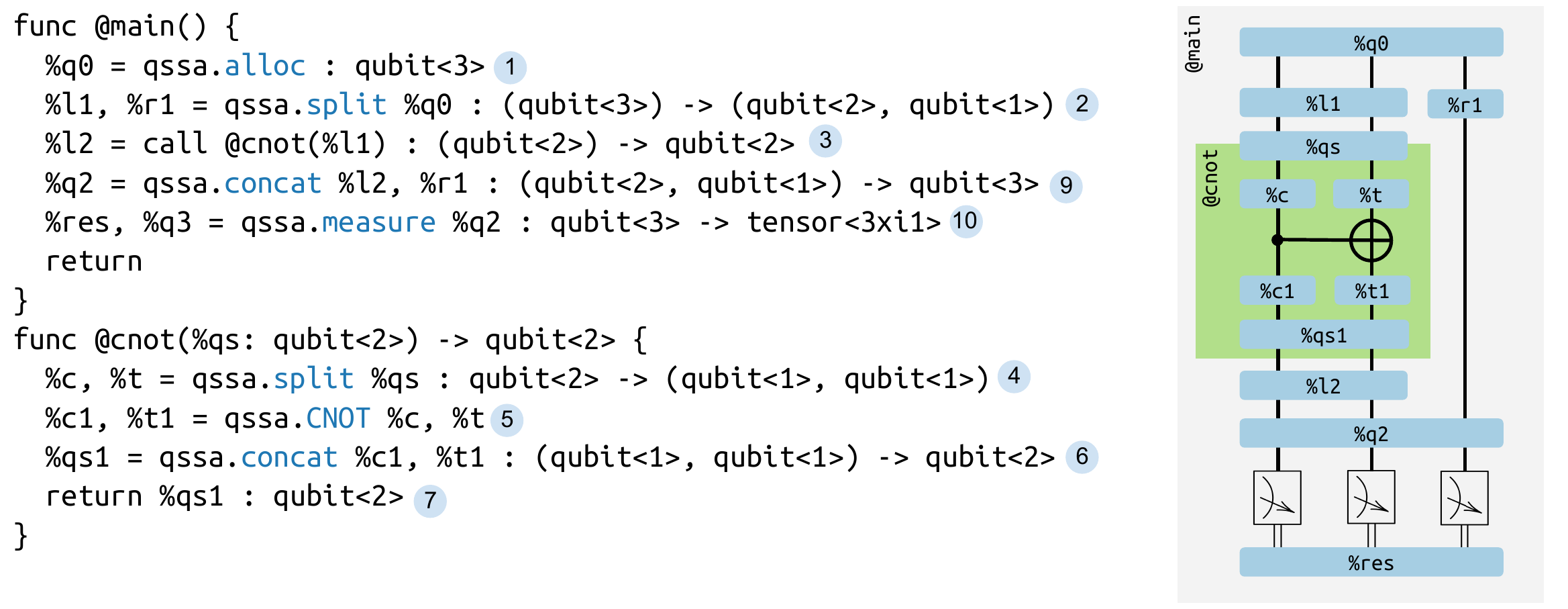}
\caption{(1) On the left is an example program showcasing core QSSA constructs.
\texttt{qssa.split}, \texttt{qssa.concat} is used to access qubits.
\texttt{qssa.CNOT} receives a qubit as input and returns a qubit as output.
Finally, measurement is performed by \texttt{qssa.measure}. (2) On the right
is the corresponding quantum circuit, and the physical qubits corresponding
to the SSA qubits. For sake of brevity, we do not show the qubits returned by
measure (\%q3)}
\label{fig:hello-world-program}
\end{figure}

\begin{figure}[!htb]
\includegraphics[width=\textwidth]{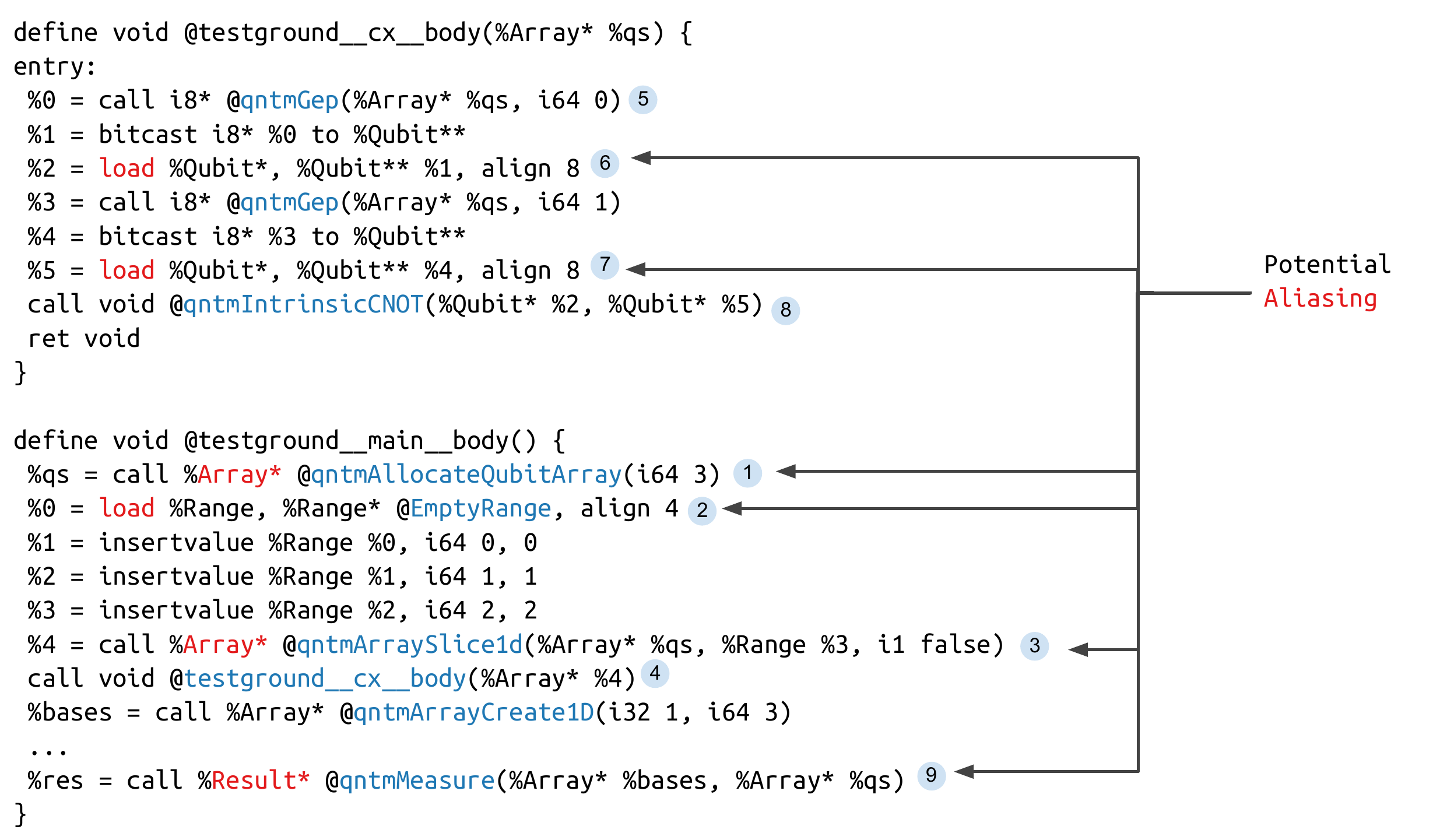}
\caption{Example program contrasting QIR with QSSA. All quantum operations are encoded using
        memory-based semantics.}
\label{fig:hello-world-program-qir}
\end{figure}

Consider the program in \autoref{fig:hello-world-program}. We first
allocate a qubit array at \circled{1} using the \texttt{qssa.allocate} instruction to
allocate an array of 3 qubits. We wish to isolate the first two qubits from the
array of three qubits to apply a Controlled-X operation on them.
To do this, we break apart the 3 element array into a (2 element, 1 element) pair
using the \texttt{qssa.split} instruction at \circled{2}.
We dispatch the qubit pair to a Controlled-X operation at \circled{3}  with \texttt{call}.
\circled{4} decomposes the two-qubit array further into a pair of single qubits,
\circled{5} invokes an inbuilt CNOT instruction, and \circled{6} puts back the pair of two single
qubits into a two-qubit array with \texttt{qssa.concat}. We return control
flow back to the main program at \circled{7}. Next, we put back the original
three qubit array with another call to \texttt{qssa.concat} at \circled{8}.
Finally, we measure the state of all three qubits with \texttt{qssa.measure}
to convert qubits into classical measurements at \circled{9}. There is an
explicit one-to-one correspondence between the program and the quantum
circuit it describes, as seen in \autoref{fig:hello-world-program}. The
circuit has two physical operations, $CNOT$ and measure, and the rest
describe the mapping between the physical qubits and the QSSA qubits arrays.

We contrast this to the QIR program in \autoref{fig:hello-world-program-qir}.
First, the array of three qubits is allocated at \circled{1} with a call to an intrinsic \texttt{qntmAllocateQubitArray}
\footnote{Names are changed for brevity. Code is left unaltered.} which returns a \emph{pointer}
to an array. Next, the \texttt{@EmptyRange} global is accessed at \circled{2} with a \texttt{load} instruction,
which manipulates global memory, modeled as a \emph{pointer load}. Next, the array is filled
with empty qubits, after which a call is made to invoke the \texttt{cx} function at \circled{3}.
Lines \circled{5}, \circled{6} first construct an offsetted \emph{pointer} with \texttt{qntmGep} and
then load the value pointed to with a \texttt{load} instruction at \circled{6}. Note that the
value loaded, \texttt{\%2}, ostensibly is a qubit and yet is modeled as a \emph{pointer to a qubit}.
Similarly, line \circled{7} also loads from a pointer. Finally, the intrinsic \texttt{qntmIntrinsicCNOT} is
invoked at \circled{8} to compute a CNOT operation. No value is returned back to \texttt{main}, since the
quantum semantics are modeled based on side-effects. Finally, at \circled{9}, we measure the qubits
with a call to the intrinsic \texttt{qntmMeasure}.

We see that throughout, QIR uses pointer based, side-effecting semantics rather
than value-based non side-effecting(pure) semantics. Since QIR is implemented
as an extension of LLVM, it implements quantum operations by hiding the
semantics of the quantum intermediate representation behind \emph{opaque
pointers}. This process, by construction, inhibits LLVM's ability to reason
about the QIR encoding, as the value semantics of modifying qubits is
complexified into the pointer semantics of manipulating an opaque pointer with a
function call. Hence, QIR is hard to analyze and optimize.

In contrast, we see that QSSA provides value semantics and pure computation,
which aids analysis and optimization, as we will see in
\autoref{sec:optimization}.  Now that we have seen a representative example of
QSSA, we shall survey the type system, instructions, and single-use constraints
of the IR.

\subsection{Type system}
QSSA has a qubit array type called \texttt{qubit<>}. The array can be either
statically sized, represented as \texttt{qubit<10>}, or dynamic sized,
represented as \texttt{qubit<?>}. We support dynamically sized arrays to
express functions that are generic over the size of the array, as described
later in \autoref{fig:qssa-array-manip-ops}.

\subsection{Classical Operations}
We use MLIR's \texttt{std} for classical arithmetic and logical operations.
We use MLIR's \texttt{scf} dialect, which provides \texttt{scf.if} and
\texttt{scf.for}, which encodes high-level control flow directly
within the intermediate representation with the use of regions.

\subsection{Quantum Operations}
QSSA has operations for (1) allocating qubits, (2) applying gates
to qubits, (3) measuring qubits to extract classical bits, (4) manipulating
qubit arrays while preserving the single-use semantics.

\subsubsection{Allocation} \texttt{qssa.alloc} allocates an
array of qubits of fixed or variable length.\\
\texttt{\%q = qssa.alloc : qubit<10>}

\subsubsection{Gate Application} We provide a complete basis set of quantum
gates\footnote{For a full reference of the gates and their matrices, refer to
\cite[Nomenclature and notation]{chuangQIQC}}, given by:
\begin{itemize}
  \item \texttt{qssa.CNOT}: The controlled NOT gate.
  \item \texttt{qssa.X}, \texttt{qssa.Y}, \texttt{qssa.Z}: The $X, Y, Z$ Pauli gates.
  \item \texttt{qssa.H}: The Hadamard gate.
  \item \texttt{qssa.Rx}, \texttt{qssa.Ry}, \texttt{qssa.Rz}: Rotation gates
      that rotate about the $X, Y$ or $Z$ axis by a given angle.
  \item \texttt{qssa.S}, \texttt{qssa.Sdg}, \texttt{qssa.T}, \texttt{qssa.Tdg}: The Phase Gate, T Gate, and their adjoints.
  \item \texttt{qssa.gate}, \texttt{qssa.U}: 
    As there are many possible choices of universal gate sets for quantum
    programs~\cite{deutsch1995universality}, we provide a generic
    instruction to define arbitrary gates. These can be used
    to define gates by specifying their matrix form.
    The $U$ gate is a convenience to describe an arbitrary
    single-qubit unitary by specifying its ZYZ Euler angles.
\end{itemize}

\subsubsection{Measurement}
The \texttt{qssa.measure} operation takes a qubit array, measures the qubits
in the standard (\texttt{Pauli-Z}) basis and returns the results as a 1D
tensor of bits (\texttt{i1}), along with the qubits.

\subsubsection{Array Manipulation}
All QSSA operations receive qubit arrays as input. To support slicing
and indexing arrays without losing our single-use constraint, we provide four primitives:
\texttt{qssa.split}, \texttt{qssa.concat}, \texttt{qssa.dim} and \texttt{qssa.cast}

\begin{itemize}
  \item \texttt{qssa.split} takes a qubit array and returns the split into
  two parts. The sizes of the two parts are indicated in the return type.
  \item \texttt{qssa.concat} takes two qubit arrays and concatenates them
  into a single qubit array.
  \item \texttt{qssa.dim} takes a qubit array and returns its size and the array itself.
  \item \texttt{qssa.cast} takes a qubit array and casts it to a different
  type. This can be used to cast a static array to a dynamic one, or vice
  versa.
\end{itemize}

\begin{figure}[!htb]
\includegraphics[width=\textwidth]{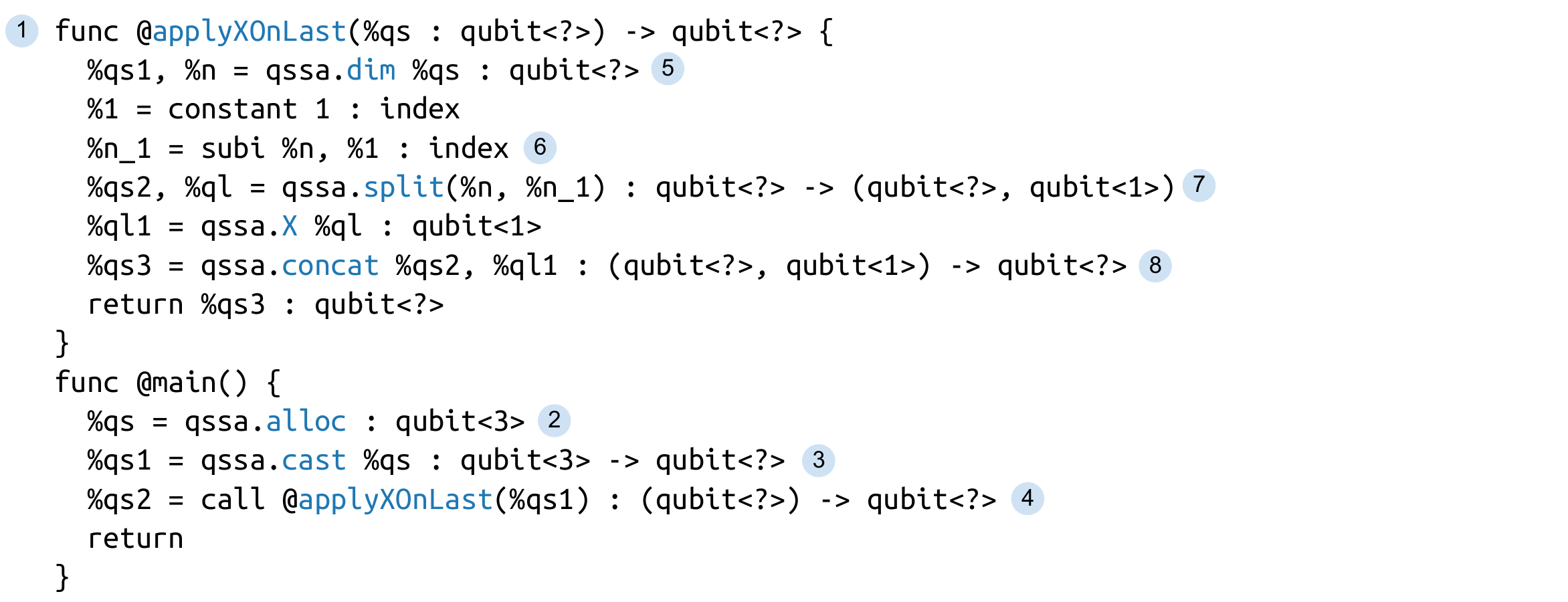}
\caption{Example QSSA program showing dynamic sized qubit arrays and the
array manipulation operations}
\label{fig:qssa-array-manip-ops}
\end{figure}

Consider the program in \autoref{fig:qssa-array-manip-ops}. The function
\texttt{@applyXOnLast} \circled{1} takes a dynamic sized qubit array and
applies \texttt{X} on the last qubit in the array. We allocate a three qubit
array \circled{2}, and cast it to a dynamic qubit array type \circled{3}. The
converted qubit array is now passed to the function at \circled{4}. In the
function \texttt{@applyXOnLast}, we first get the dimension of the argument
\circled{5}. We then subtract one from it at \circled{6}, to compute the size
of the left side array. We then call \texttt{split} at \circled{7}, passing
$n$ and $n - 1$ to it, which correspond to the sizes of each of
the dynamic arrays in the type signature. We then apply \texttt{X} on the
last qubit, and concatenate them back at \circled{8}.

\subsection{Semantics: Side-effect free operations}
All QSSA operations (except \texttt{alloc}) have no side-effects. They take
input qubits, operate on them and return the resulting qubits - which
effectively correspond to the same physical qubits in the underlying system.
We explain how this enables us to reuse SSA optimization in
\autoref{subsec:reusing-ssa-optimizations}

\subsection{Single Use Analysis}

Each SSA value representing a qubit must be used only once to enforce
the no-cloning constraint. We describe an analysis for an SSA-based IR with CFGs
that enforces this constraint. We then adapt this analysis to MLIR, which offers
nested CFGs using regions.

\begin{figure}[htb]
    \includegraphics[width=\textwidth]{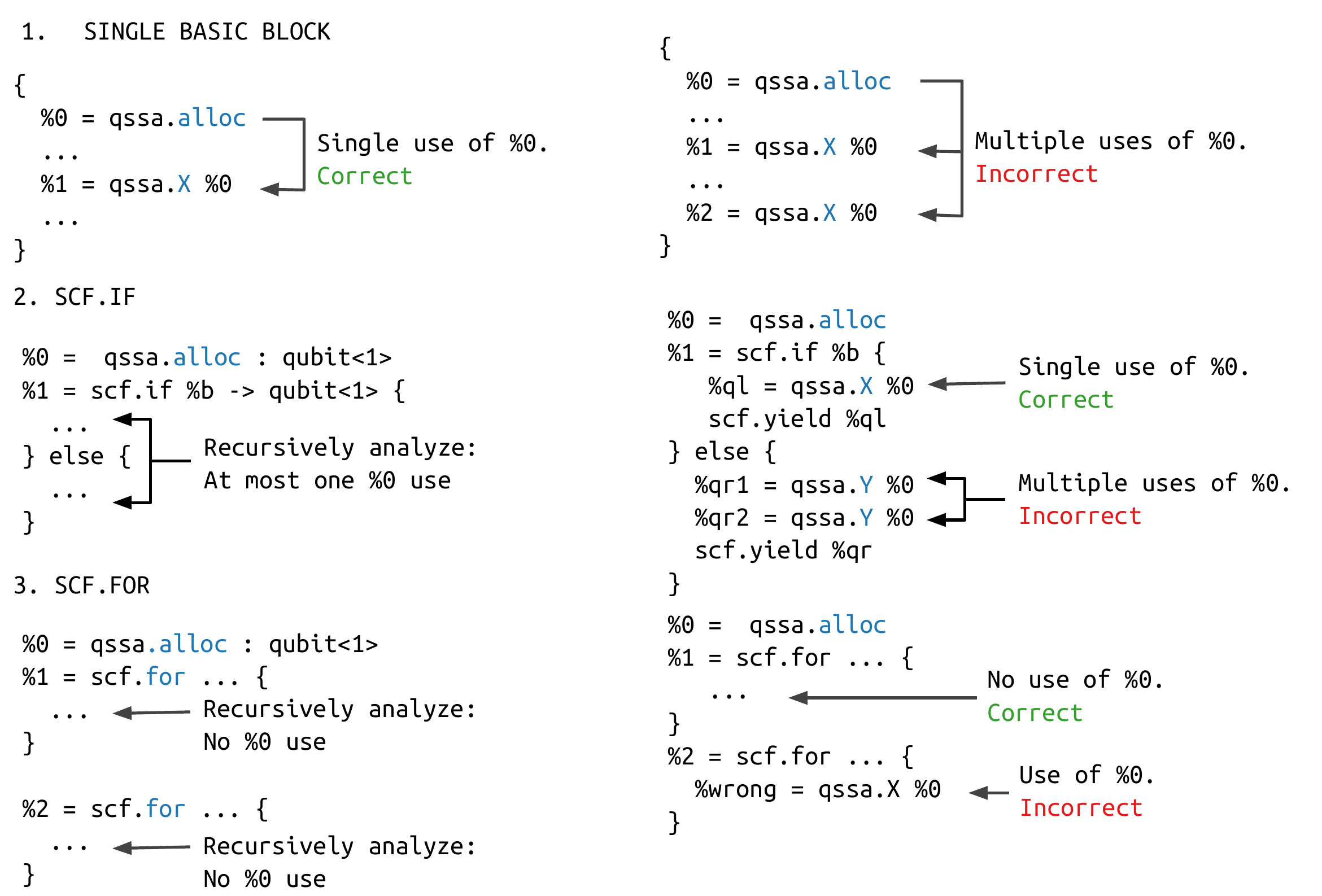}
    \caption{Linear time algorithm to determine qubit single-use in a program with straight line code,
        and control flow represented using \texttt{scf.if}, \texttt{scf.for}}
\end{figure}

\subsubsection{Algorithm for flat CFG}
In this scenario, we consider a CFG with no cycles.  For each qubit SSA value
\texttt{\%q}, consider all uses. (1) If a use is within the same basic block,
then the qubit is dead, and we check that there is no other use in any other
basic block. (2) If there are uses in other basic blocks, then for every pair of
uses $u_i$ and $u_j$, we check that there is no path from $u_i$ to $u_j$ and
vice versa. This can be done in linear time using dynamic programming on the CFG
DAG. We process the blocks in reverse topological order. If the block has a use
of the qubit, we mark the block as used. If any child block has been marked,
then we mark the current block as well. If the block gets marked twice, then
there is a repeated use.

\subsubsection{Adaptation to Regions}

Regions can be used to represent higher-level control flow constructs.
Representing control flow such as \texttt{if} conditions and \texttt{while
loops} directly in the IR allows us to infer facts about control flow without
building and querying the dominator tree, as we would have to for an arbitrary
CFG. We exploit this to optimize our single-use analysis by assuming a regular
program structure that consists of only a single basic block, with all control
flow represented by nested regions using \texttt{scf.if} (if-then-else), and
\texttt{scf.for} (for loops). This allows us to deduce a \emph{linear time}
algorithm to verify single-use.

\paragraph{Constraints} We must verify the following constraints:
(1) A qubit can be used at most once in the same region. 
(2) If the qubit has two uses in different regions, then neither region must be
an ancestor of the other.
(3) If a qubit is used in a region which has an \texttt{scf.for} as an ancestor,
then the region of its definition must also be a descendent of the same for
loop.

\paragraph{Algorithm} We consider the topmost region (could be a function body)
which is isolated from above. We perform a depth first search from this root
region on the region tree induced by the control flow operation. We maintain two
sets of qubits: ones that are defined but not used $D$, and ones that are used -
$U$. 
At each region $R$, we perform the following:
\begin{enumerate}
  \item Maintain a set $B$ for rolling back updates.
  \item Iterate through the operations linearly. For each operation, we verify
  that all its qubit arguments $q$ lie in $D$. If so, we move $q$ from $D$ to
  $U$, and add $q$ to $B$.
  \item For all the \texttt{scf.if} operations, recursively verify the
  constraints, by passing the sets $D' = D, U' = U$.
  \item For all the \texttt{scf.for} operations, recursively verify the
  constraints, by passing the sets $D' = \texttt{iter\_args}, U' = \{\}$.
  \item Rollback all the qubits in $B$, by moving them from $U$ to $D$.
\end{enumerate}

\section{QSSA Optimizations}
\label{sec:optimization}

QSSA allows us to easily adapt many classical SSA analyses and optimizations to
the quantum regime.  Here, we outline both our single-use analysis, as well as
adaptations of classical SSA transforms such as common sub-expression
elimination. 

\begin{figure}[htb]
    \includegraphics[width=\textwidth]{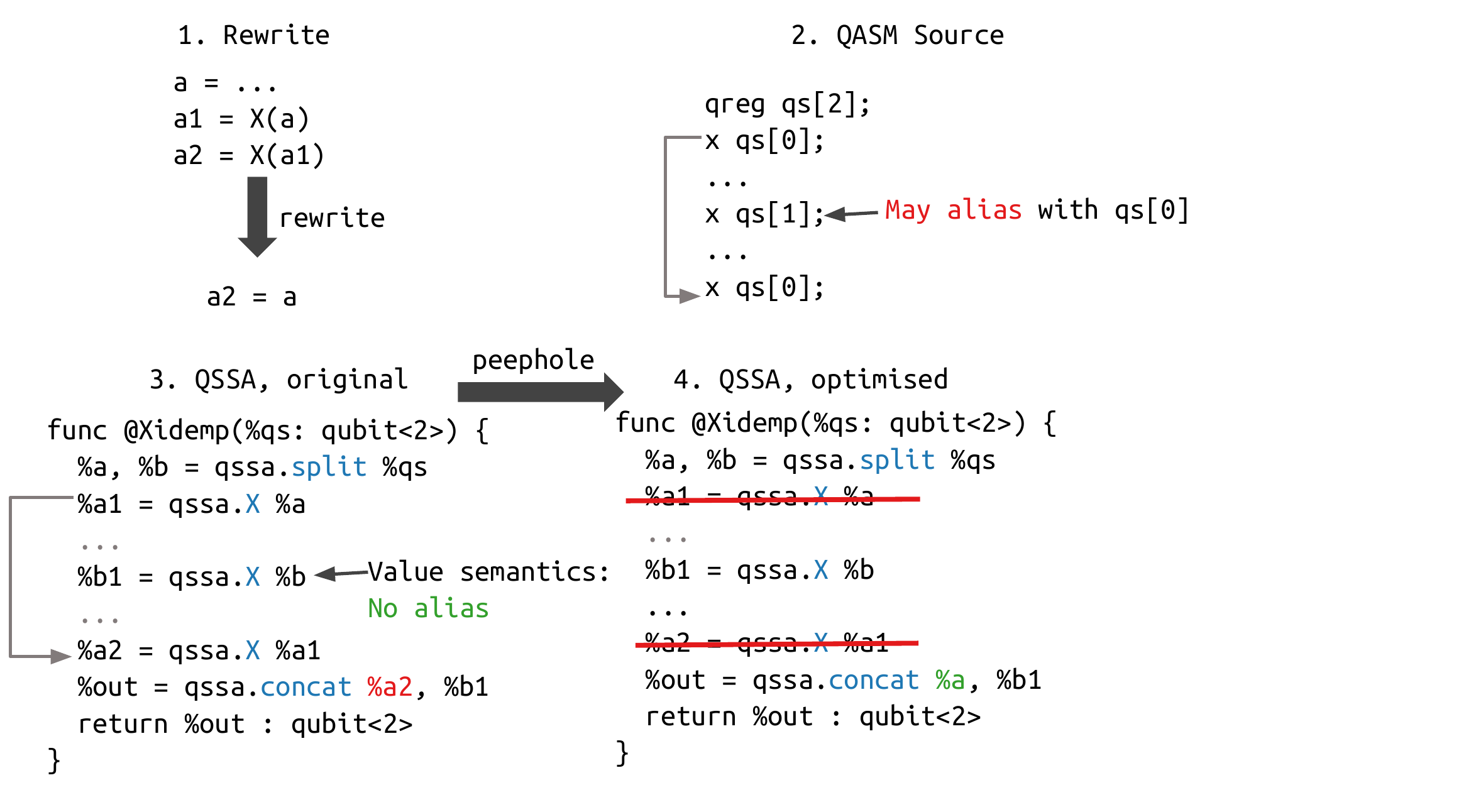}
    \caption{Example optimization using the identity that the $X$ gate is a
    self inverse. (1) describes the rewrite rule. (2) shows the roadblock in
    rewriting the pattern in OpenQASM. (3), (4) show the rewrite in QSSA.
    Since SSA carries data dependencies explicitly, one can pattern-match on
    the IR, rather than having to reason about array aliasing.}
             \label{fig:peephole-x-cancel}
\end{figure}
\subsection{Peephole Optimizations}
\label{subsec:peephole}
QSSA provides classical peephole optimizations on the
quantum circuit, where we look for sub-circuit patterns, and then replace
matches with a more optimized version of the sub-circuit being matched.
The purity of our IR enables us to pattern-match and rewrite directly on the SSA DAG,
since we can ignore the ordering of instructions as there are no side
effects. An example transformation in \autoref{fig:peephole-x-cancel} we
perform, the Pauli X gate is a self-inverse. Thus, applying the Pauli gate
twice can be optimized away into a no-op. This can be concisely stated using
the MLIR pattern matching framework, as the data dependencies are encoded.

In total, we implement the following matrix identities:
\begin{enumerate}
    \item CNOT-CNOT optimization.
    \item $U$ gate merging.
    \item Pauli relations: $X^2 = Y^2 = Z^2 = -iXYZ = I$. 
    \item Hadamard self-adjoint: $H^2 = I$
    \item Drop phase before measurement.
\end{enumerate}

\begin{figure}[htb]
    \includegraphics[width=\textwidth]{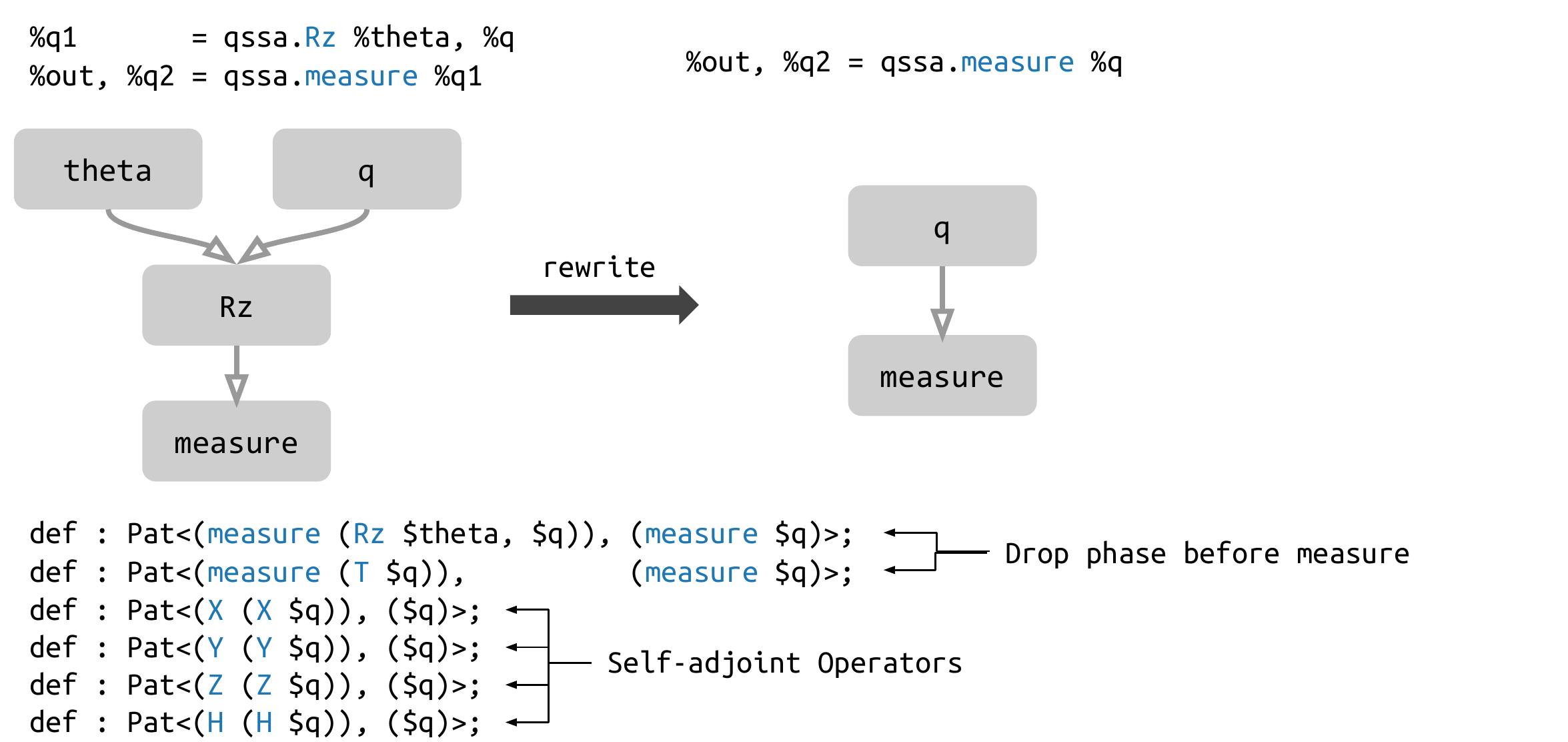}
    \caption{Declarative encoding of peephole optimizations. The abstract DAG transform is shown
             above, which eliminates the \texttt{Rz} gate immediately before a measure. The 
             set of rewrites which eliminate gates before measurement is shown
             below. The encoding is written in TableGen, an declarative SSA
             pattern matching infrastructure for MLIR.}
             \label{fig:decl-rewrites-tablegen}
\end{figure}

Our implementation of these peephole optimizations is declarative. This is in
contrast to Qiskit, where the optimization passes are written in imperative
Python code that manipulates internal data structures. We use MLIR's TableGen
infrastructure to declare DAG-to-DAG rewrites, which are then automatically
applied during canonicalization. \autoref{fig:decl-rewrites-tablegen}
showcases a few examples of the above mentioned rewrite patterns.

\subsection{Redundancy Elimination}
\label{subsec:redundancy-elimination}
QIR and QSSA model qubits with side-effecting memory operations and how this
inhibits optimizing code where qubits may alias.  In contrast, QSSA trivially
exploits SSA's redundancy elimination to optimize examples that QIR is unable
to.

\begin{figure}[htb]
  \includegraphics[width=\textwidth]{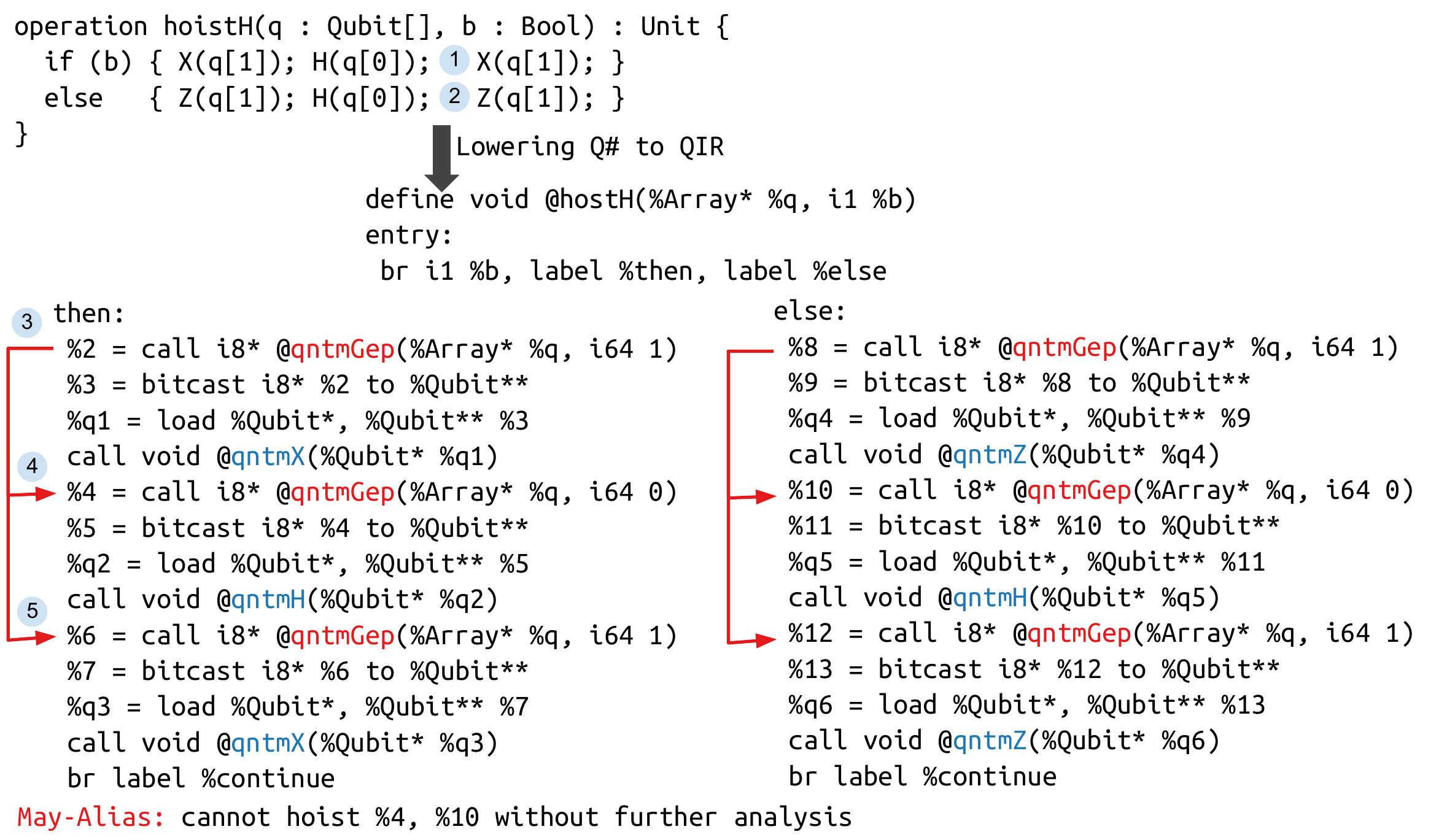}
  \caption{A QIR program where the compiler is unable to hoist the call
    \texttt{H(q[0])} due to potential aliasing of array indexes in the IR.
     The LLVM \texttt{getelementptr} instruction is adapted to \texttt{qntmGep},
     an opaque intrinsic whose semantics LLVM does not know. Similarly, the
     types \texttt{Array} and \texttt{Qubit} are also opaque LLVM types, whose contents
     and semantics are unknown to LLVM by design. QIR does not interoperate
     with the LLVM infrastructure, making alias and side-effect analysis
     impossible, thereby preventing any SSA-based optimization from being performed.}
  \label{fig:qir-redundancy-elimination}
\end{figure}

\begin{figure}[htb]
  \includegraphics[width=\textwidth]{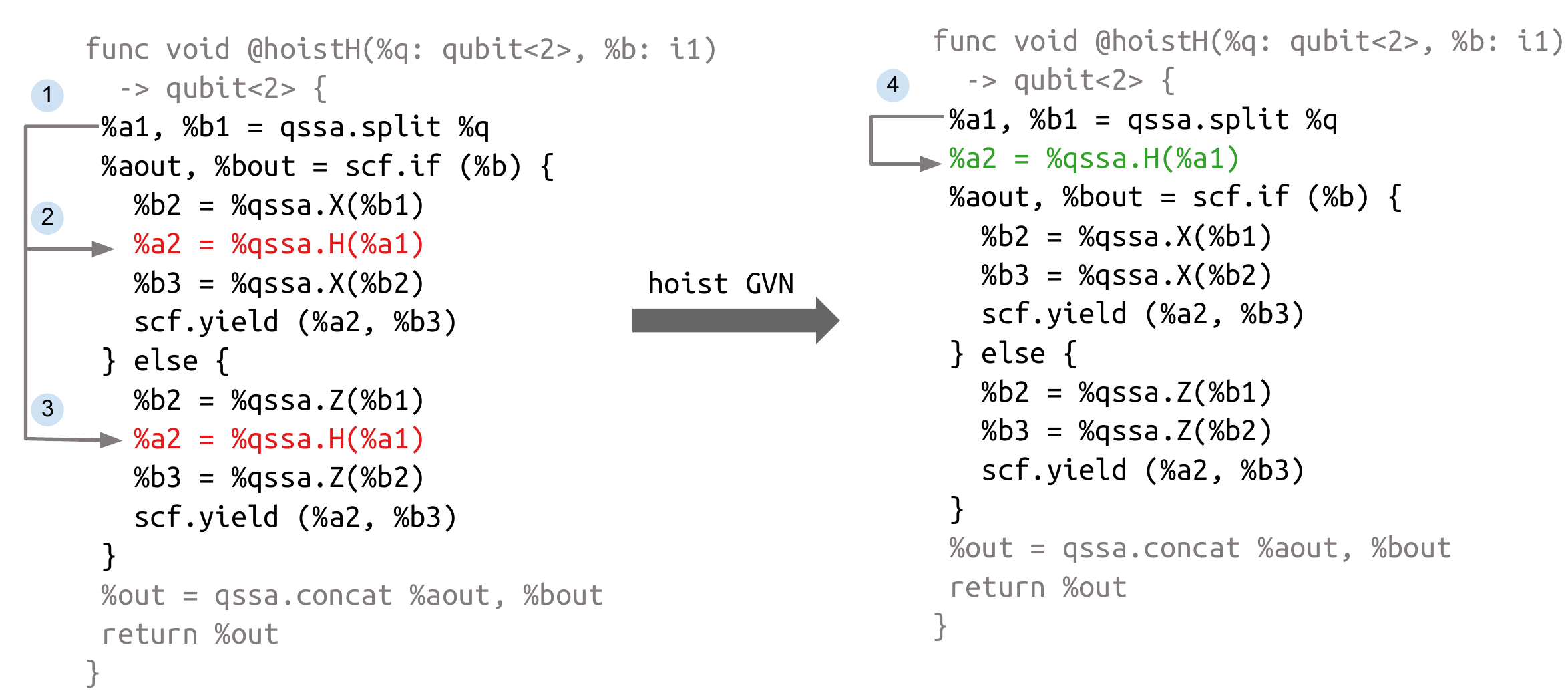}
    \caption{A QSSA program that exhibits redundancy elimination. QSSA
    expresses data dependencies in the use-def chain, and has no side effects,
    thus allowing global value numbering to hoist the instruction out of the
    branch.}
  \label{fig:qssa-redundancy-elimination}
\end{figure}

In \autoref{fig:qir-redundancy-elimination}, we consider redundancy elimination
for QIR.  We should be able to hoist the operation \texttt{H(q[0])} in the \texttt{then}
and the \texttt{else} branch, at locations \circled{1} and \circled{2} since it
occurs in both branches of the Q\# source code. The generated QIR
occludes the transformation by using array indexing
semantics with \texttt{qntmGep} at \circled{3}, \circled{4}, \circled{5} in the \texttt{then} branch,
and similarly in the \texttt{else} branch. Hoisting the \texttt{H(q[0])} instruction
encoded by \texttt{call void @qntmH(\%Qubit* \%q2)} requires alias analysis to ensure
that the qubit \texttt{\%q2} does not alias with \texttt{\%q1, \%q3}.
QIR is unable to perform this alias analysis and is thus unable 
to optimize this program \footnote{We generate QIR from Q\#, and then optimize
using \texttt{opt -O2}, on Q\# version \texttt{Microsoft.Quantum.Sdk/0.15.2103133969}, LLVM version \texttt{11.1.0}}. In contrast, the QSSA pipeline shown in \autoref{fig:qssa-redundancy-elimination}
explicitly encodes qubit dependencies in the def-use chain. It is clear that the \texttt{qssa.H}
instructions have the dependency chain \circled{2} $(\texttt{a2} \mapsto \texttt{a1})$ \circled{1}.
The hoisting analysis ignores the interleaved \texttt{\%b2 = \%qssa.X(\%b1)}
instruction. This allows QSSA to hoist the instruction out of the
\texttt{scf.if}, as shown at \circled{4}. This optimization is performed
\emph{entirely} by the inbuilt global value numbering pass, aided by our SSA
encoding of quantum semantics.

\subsection{Reusing SSA optimizations}
\label{subsec:reusing-ssa-optimizations}
All quantum operations except \texttt{alloc} have no side-effects. That is,
the results of such an operation is completely determined by its inputs, and
it does not depend on any external state, nor does it change it. The Allocate
operation \texttt{alloc} has a minor side-effect just like a memory allocate
instruction (eg. \texttt{malloc}). It acquires a resource but does not
modify it. We only allow optimization passes that do not create copies of SSA
values, since these preserve no cloning. Thus, we reuse:

\begin{itemize}
        \item Inlining
        \item Loop Unrolling
        \item Dead code elimination
        \item Common sub-expression elimination
\end{itemize}

\section{Raising and Lowering}
In this section, we describe how our QSSA interacts with non-SSA languages.
We describe how we \emph{raise} OpenQASM, a low-level non-SSA-based quantum
IR into QSSA for analysis and optimization. We then describe how to
\emph{lower} QSSA code to OpenQASM.

\subsection{Raising OpenQASM to QSSA}
To raise from OpenQASM to the QSSA IR, we perform an analogue of
mem2reg to convert memory references into SSA def-use
chains. To make the conversion simpler, we provide a \texttt{qasm} dialect in
MLIR, which is a 1-1 mapping of the OpenQASM operations. The OpenQASM code is
first converted to MLIR, using the quantum operations of the \texttt{qasm}
dialect, and the classical operations of the \texttt{std} and \texttt{scf}
dialects.

\begin{figure}[htb]
\includegraphics[width=\textwidth]{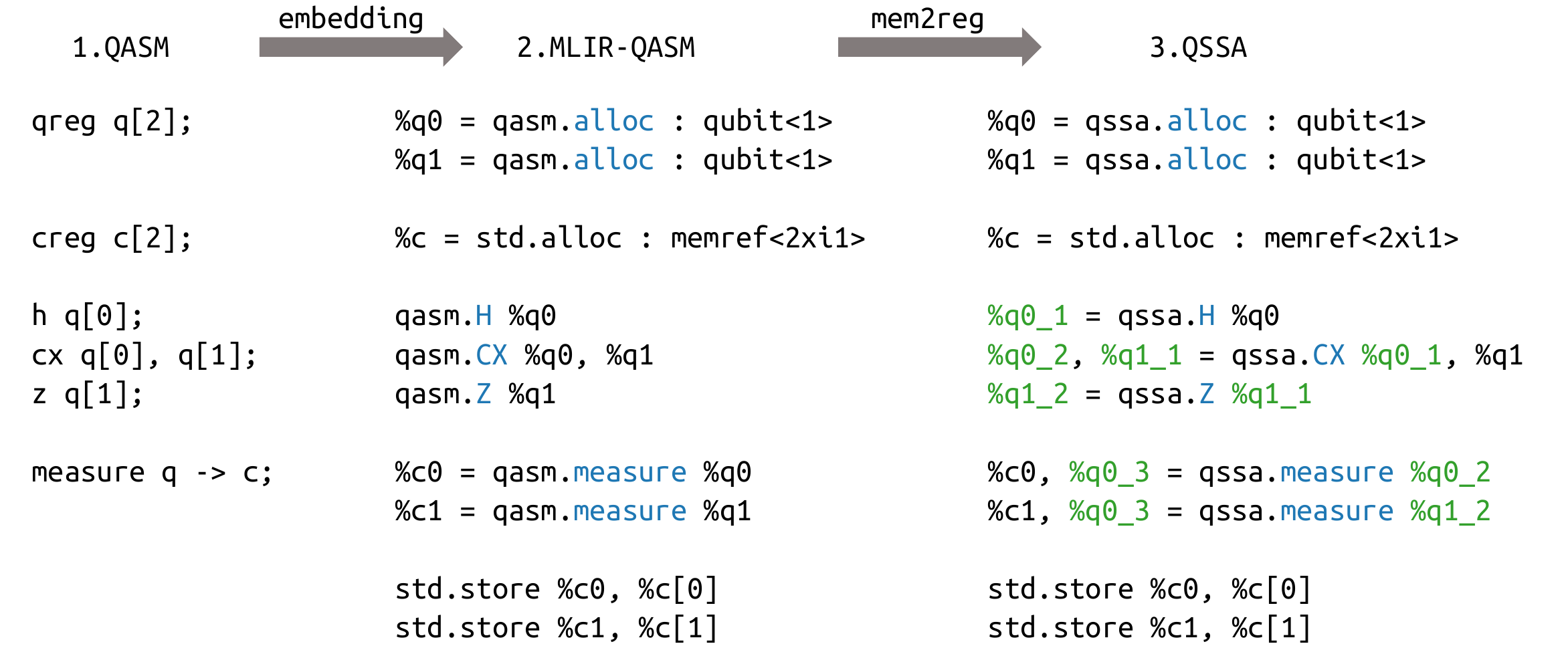}
\caption{Sample program showing conversion from OpenQASM to QSSA. We first embed OpenQASM
         within MLIR, after which we perform a \texttt{mem2reg}-like pass to raise
         qubit memory operations into SSA. New values that are inserted
         during \texttt{mem2reg} are highlighted in green.}
\label{qasm-to-qssa-example}
\end{figure}

\subsubsection{Latest Qubit Map}
We maintain a map data-structure \texttt{M} that maps each OpenQASM qubit to
its latest SSA result that refers to this qubit. When an operation is applied
on this qubit, we update the map with the new result. This enables us to do
the mem2reg transformation.

\subsubsection{Quantum Registers}
As OpenQASM only allows statically sized arrays, we simplify the conversion
by allocating single qubits - instead of allocating a qubit array and then
splitting. This removes unnecessary \texttt{split} and \texttt{concat}
operations. In \autoref{qasm-to-qssa-example}, we convert the two qubit
register \texttt{q[2]} to two \texttt{qasm.alloc} operations. We also
initialize \texttt{M[\%q0] := \%q0} and \texttt{M[\%q1] := \%q1}.

\subsubsection{Gates}
To convert a \texttt{qasm} gate, we first extract the latest SSA qubits for
each of the gate's operands. We emit an \texttt{qssa} gate operation with
those SSA qubits as inputs, and update the map with the results. In
\autoref{qasm-to-qssa-example} to convert the \texttt{qasm.H \%q0} operation,
we emit \texttt{\%q0\_1 = qssa.H \%q0} and update the qubit map
\texttt{M[\%q0] := \%q0\_1}. (Note that \texttt{\%q0\_1} corresponds to
physical qubit \texttt{q[0]} in the OpenQASM code.)

\subsubsection{Measure}
In the MLIR \texttt{qasm} dialect, we decouple the measurement and store
operations. \texttt{qasm.measure} takes a qubit and returns a bit, and then
it is stored in the \texttt{memref} by an explicit \texttt{store} operation.
We convert this to QSSA similar to converting gates - measure operation is
syntactically like a gate, which returns an extra bit (along with the input
qubits).

\subsection{Lowering QSSA to OpenQASM}
To lower, we first convert all operations in the \texttt{qssa} dialect to the
\texttt{qasm} dialect. Then we directly translate all \texttt{qasm}
operations to OpenQASM operations.

\begin{figure}[htb]
\includegraphics[width=\textwidth]{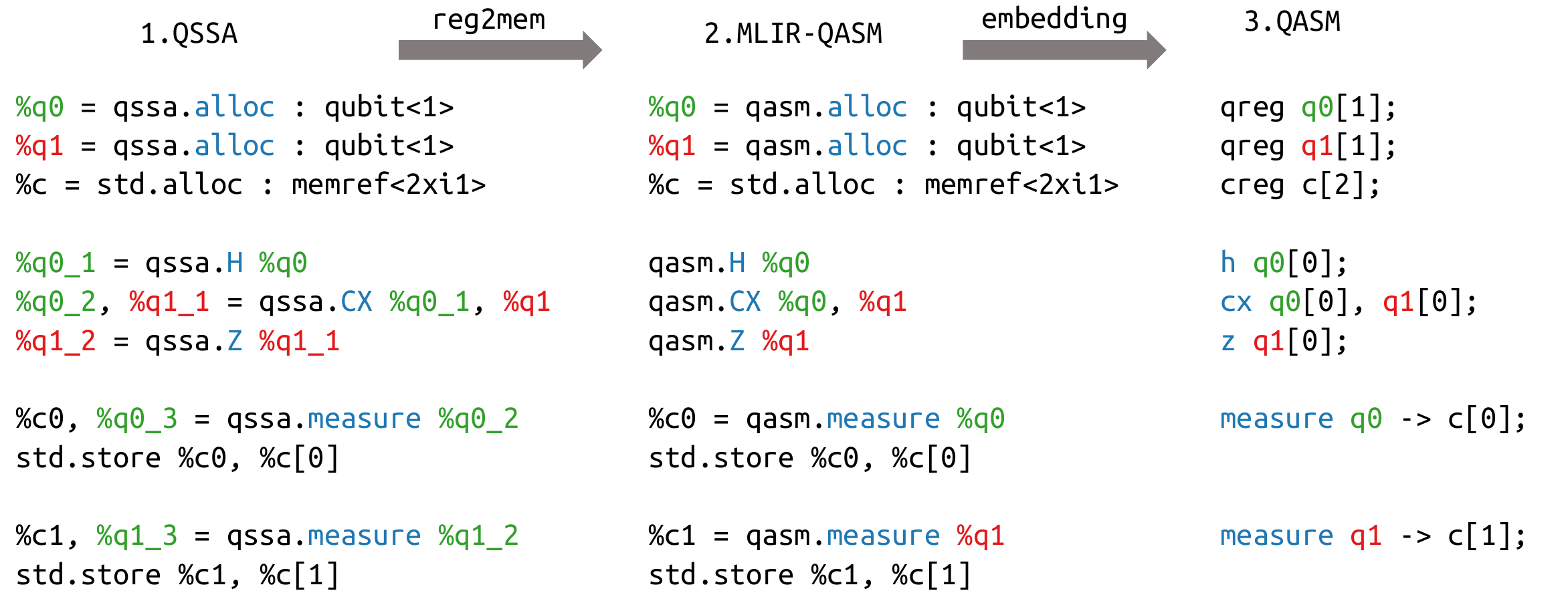}
\caption{Sample program showcasing conversion from QSSA to OpenQASM. We first lower from QSSA
    to MLIR-QASM by eliminating qubit registers. We then translate MLIR-QASM into QASM. The
    two physical qubits in the program are color coded as light green and dark red respectively.}
\label{fig:qssa-to-qasm-example}
\end{figure}

\subsubsection{Allocation} \texttt{qssa.alloc} operations are directly converted
to \texttt{qasm.alloc}. Classical registers are identified by \texttt{memref}
of type \texttt{i1}. We declare \texttt{creg}s for each of those
\texttt{memref.alloc} operations.

\subsubsection{Gates}
A gate operation is converted by removing its return values. Instead, we
replace its current return values with its operands in the rest of the IR. In
\autoref{fig:qssa-to-qasm-example}, we convert the \texttt{\%q0\_1 = qssa.H
\%q0} instruction to \texttt{qasm.H \%q0}. And then we replace the result
\texttt{\%q0\_1} with the operand \texttt{\%q0} - at the \texttt{qssa.CX}.

\subsubsection{Measurement}
We convert measure operations similarly. As measurements only return the
measurement outcome, to convert to OpenQASM, we find which qubit's measurement
result is stored in the memref, and emit the corresponding operation.

\section{Evaluation}
In this section, we evaluate our prototype against Qiskit, IBM's quantum
computing infrastructure by measuring optimizations in terms of gate count and
circuit depth reduction on two datasets: QASMBench~\cite{li2020qasmbench} and
IBM Quantum Challenge Dataset~\cite{ibmqxchallenge}.

QSSA has a set of generic optimizations derived from SSA, along with peephole
optimizations described in \autoref{subsec:peephole}. We compare our
prototype implementation against Qiskit's standard
optimizer~\cite{qiskit_transpiler_api}, and compute percentage reduction in
gate counts from the baseline. We run Qiskit at three levels of optimization
- \texttt{O1}, \texttt{O2}, \texttt{O3}.\footnote{Qiskit supports another
level for optimization - level 0 - but we omit it as it does no
optimization.} We compare them using two metrics: (1) percentage number of
gates reduced and (2) compilation/optimization time.

\subsection{QASMBench}

QASMbench is a low-level QASM benchmark suite for NISQ evaluation and
simulation. It contains 60 programs split into three categories: small (30),
medium (16), large (14).\footnote{We remove two extremely long programs which
contain >2M lines each.}. It collects common quantum algorithms and routines
from a variety of domains including chemistry, simulation, linear algebra,
searching, optimization, quantum arithmetic, machine learning, fault tolerance,
and cryptography.

  \begin{figure}[htb]
  \includegraphics[width=\textwidth]{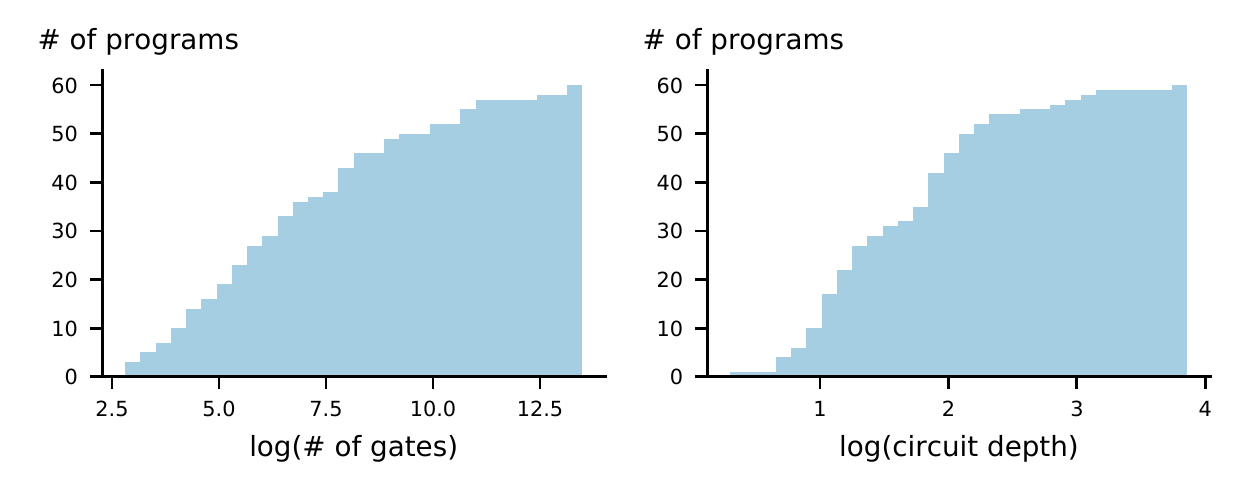}
  \caption{Cumulative histogram of logarithm of gate counts and circuit depth for the
  QASMBench dataset. We see that the dataset is well distributed, containing
  small to medium size programs }
  \label{fig:qasmbench-gate-count-gate-depth}
  \end{figure}
  \begin{figure}[htb]
  \includegraphics[width=\textwidth]{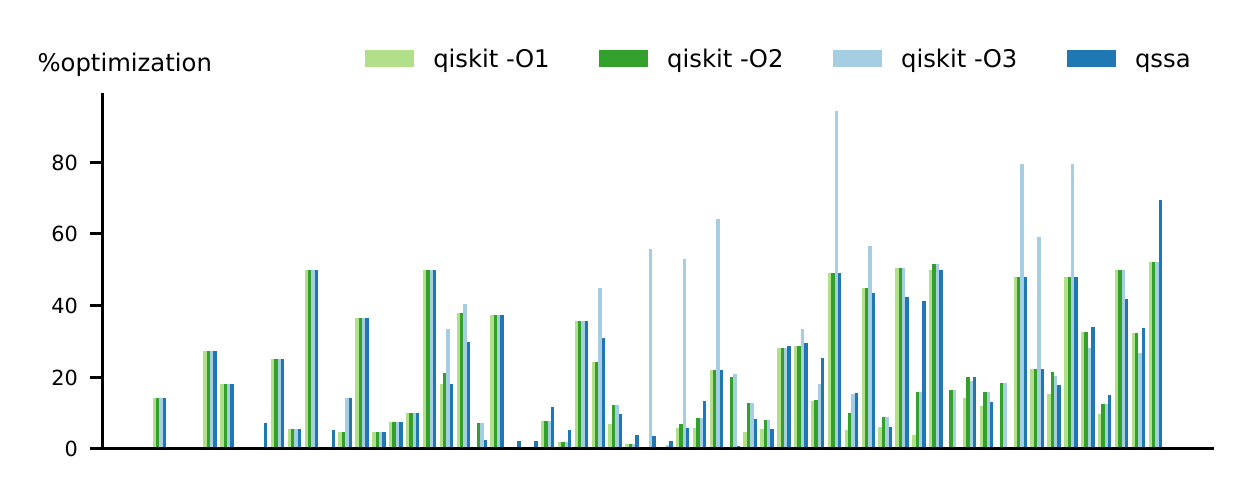}
\caption{Optimization ratios of the QASMBench dataset, calculated as the proportion of instructions
removed: $1-i_f/i_0$ where $i_0$ is the initial gate count, and $i_f$ is the
final gate count. The $x$ axis is in the ascending order of program length.
We see that we are able to optimize as well as or better than Qiskit in
majority of the cases.}
  \label{fig:qasmbench-results}
  \end{figure}
  \begin{figure}[htb]
  \includegraphics[width=\textwidth]{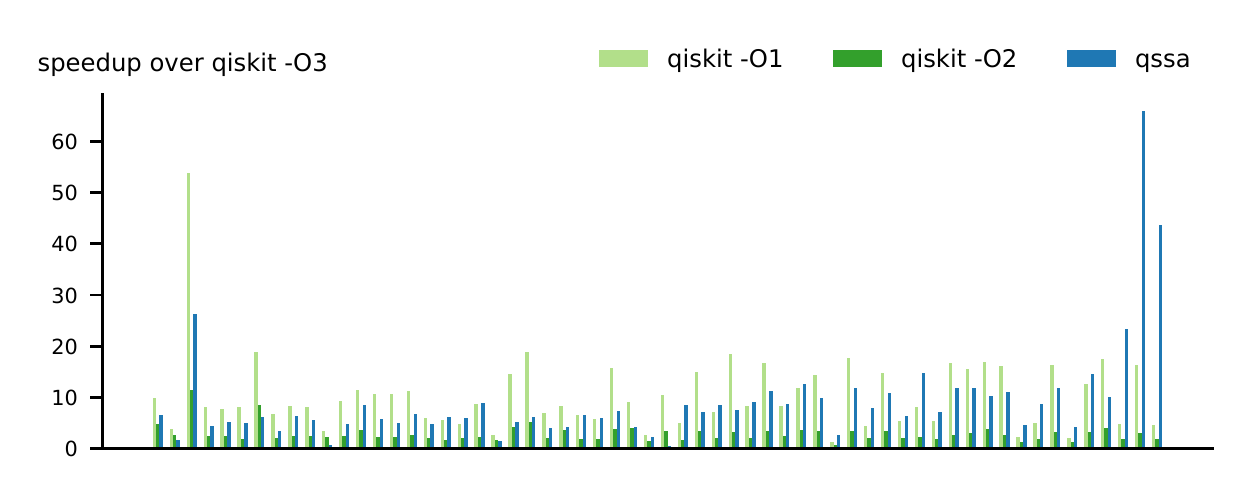}
\caption{Compilation/optimization speedup of the QASMBench dataset. The $x$ axis is in the
ascending order of program length. QSSA is much faster than O3, and comparable
to O2 and O1 in most cases.}
  \label{fig:qasmbench-times}
  \end{figure}


\autoref{fig:qasmbench-gate-count-gate-depth} plots a cumulative histogram of
the logarithm of gate count and circuit depth of the dataset. The average gate
count is 780 and average circuit depth is 252. The dataset is quite extensive,
as it covers the number of qubits ranging from 2 to 60K, and the circuit depth
from 4 to 12M.

We run the optimizers and plot the final gate counts in
\autoref{fig:qasmbench-results}. We observe that we optimize better than
Qiskit level 1 in 30 cases (50\%) , and achieve parity in 56 out of 60 cases
(93\%). Compared to Qiskit level 2, we optimize better in 21 cases (35\%) and
achieve parity in 43 cases (72\%). Compared to Qiskit level 3, we optimize
better in 18 cases (30\%) and achieve parity in 35 cases (58\%).

\autoref{fig:qasmbench-times} shows the corresponding speedups in compilation
times when compared to Qiskit \texttt{O3}. QSSA performs comparably with
\texttt{O1}, with a maximum of 2x slowdown in most cases. It also has a 2-3x
speedup over \texttt{O2} in most cases. In some extreme cases - medium
\emph{VQE} and large \emph{Ising Model}, QSSA performs extremely well with
huge speedups over Qiskit - 65x and 44x respectively against \texttt{O3} and
21x each against \texttt{O2}.

\subsection{IBM Quantum Challenge Dataset}
The IBM Quantum Challenge Dataset consists of 141 quantum programs, designed
to be complex programs to stress test quantum optimization and quantum qubit
routing algorithms.\footnote{We remove extremely large programs from the
dataset, which being synthetically generated have a size of over $10^6$
lines.}.

\begin{figure}
    \includegraphics[width=\textwidth]{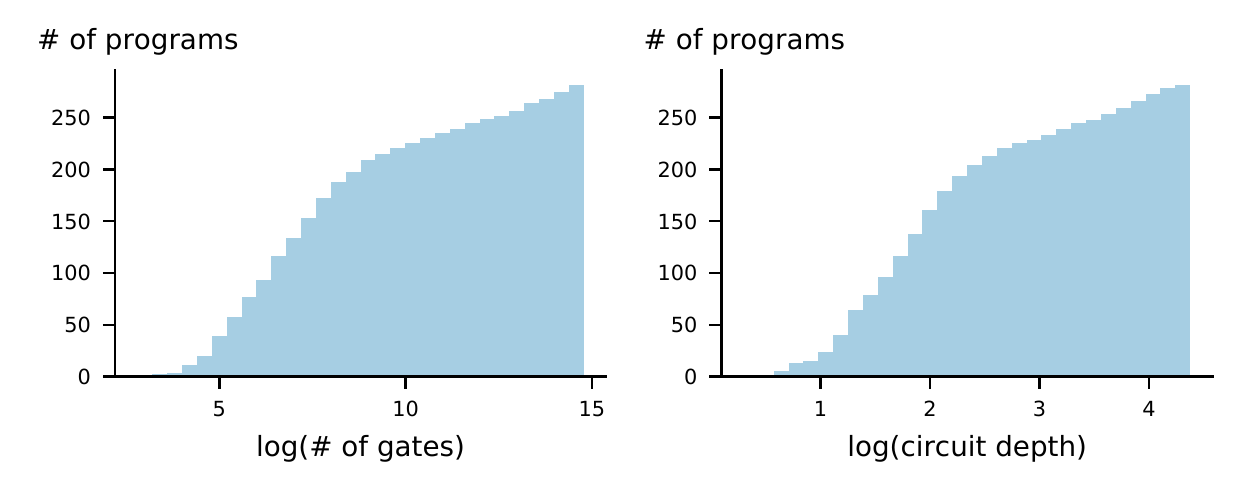}
\caption{Cumulative histogram of gate depth and gate counts for the IBM
    Challenge dataset. We see that the set of programs is quite exhaustive,
    containing both small and large programs. }
\label{fig:gate-count-gate-depth}
\end{figure}

We begin by characterizing the dataset, and we find that the
average gate count is 2198, and average gate depth is 1185. This indicates that
our programs are quite large. \autoref{fig:gate-count-gate-depth} plots a 
histogram of the \emph{logarithm} of the gate count and gate depth of the dataset.
We see that there are quite a few large programs, with gate depths of $10^4$.

\begin{figure}
\includegraphics{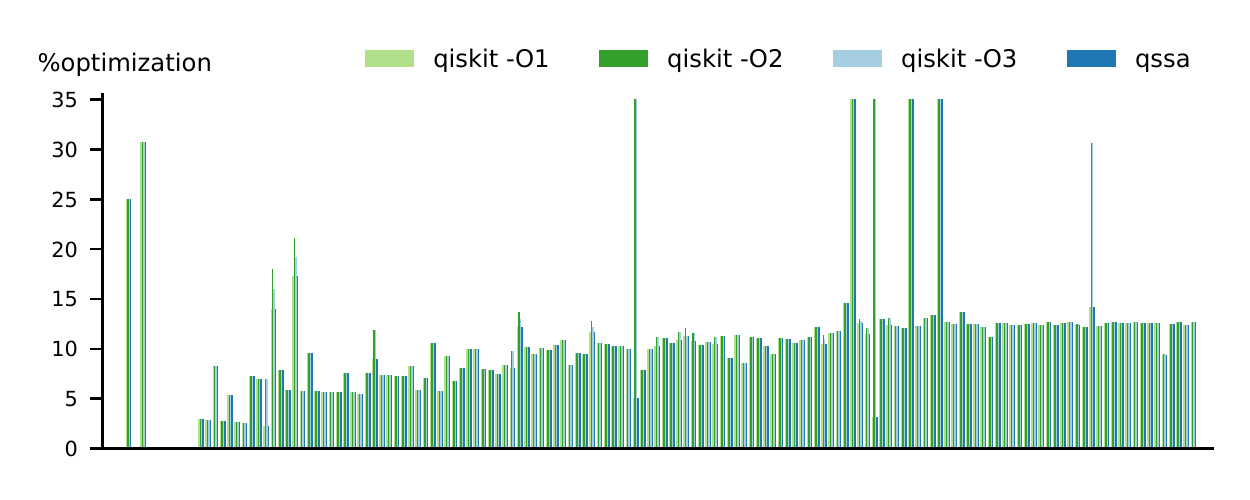}
\caption{Optimization ratios of the IBM Challenge dataset, which are calculated as the proportion of
instructions removed: $1-i_f/i_0$ where $i_0$ is the initial gate count, and
$i_f$ is the final gate count. The $x$ axis is in the ascending order of
program length. The $y$ axis is capped at 35\% for better visualization. 5
test files exceed the 35\% ratio.}
\label{fig:ibm-optimizations}
\end{figure}

\autoref{fig:ibm-optimizations} shows the optimization ratios of the dataset.
5 out of 149 programs get heavily optimized out by Qiskit \texttt{O3}, with
optimization ratios between 50\%-95\%. The three \emph{Ising Model} programs
get optimized to about 50\%, and the two \emph{QFT} programs get optimized to
89\% and 95\% respectively. We cap the Y-axis at 35\% in the plot to improve
visualization. QSSA achieves parity in all test cases when compared with
Qiskit \texttt{O1}. It achieve parity with 125 out of 149 cases (83\%)
compared against both \texttt{O2} and \texttt{O3}. We are performance
comparable with the Qiskit optimizer, without any sophisiticated rewrites
in the QSSA system. We rely purely on peephole rewrites along with
standard SSA optimizations. The Qiskit compilation suite is mature,
and QSSA being performance comparable showcases the strengths of an
SSA-based encoding.

\begin{figure}
\includegraphics{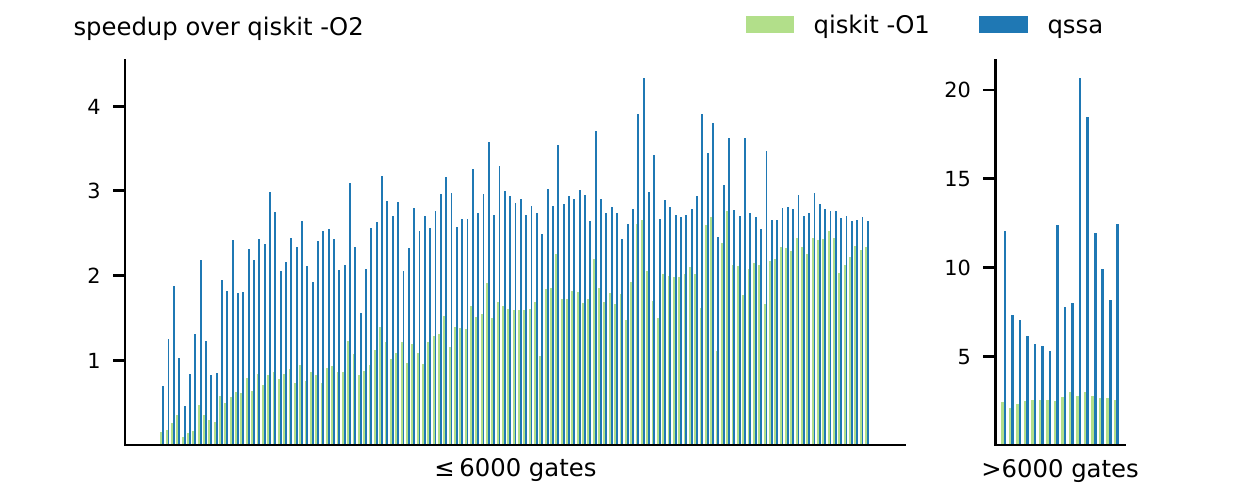}
\caption{Compilation/optimization speedup of the IBM Challenge dataset. The
$x$ axis is in the ascending order of program length. QSSA is faster than
\texttt{O1} and \texttt{O2} in almost all cases.}
\label{fig:ibm-times}
\end{figure}

We plot compilation/optimization time speedups in \autoref{fig:ibm-times}. We
split the dataset into two parts: one with small programs ($\le 6000$ gates,
133 files) and one with large programs ($> 6000$ gates, 16 files). We
consider the speedup over \texttt{O2} and plot \texttt{qssa} against
\texttt{O1} as it is the extreme (fastest) case. There is an anomaly in
really small tests, where \texttt{O2} is faster than \texttt{O1}. We conjecture
that this is due to \texttt{O2}'s awareness of certain patterns which occur
in small test cases that enable it to rapidly optimise the test case. (1) In the small set,
\texttt{qssa} is faster than \texttt{O1} in all programs, and faster than
\texttt{O2} in 96\% of them. It also has 2x speedup over \texttt{O1} in 44\%
of the programs and over \texttt{O2} in 88\% of the programs. (2) In the
large set, \texttt{qssa} is 2x faster than all programs over both Qiskit
levels. Moreover, it is 5x faster than \texttt{O2} for all programs. It is
also 3x faster than \texttt{O1} in 62.5\% of the programs. In summary, we find
that our QSSA optimizer prototyped in MLIR is competitive both in terms of
optimizing gate count and compilation time.




\section{Related Work}
The discovery of efficient quantum algorithms by Shor and Grover led to a
proliferation of quantum programming languages~\cite{miszczak2010models}.
\citet{cross2017open} introduced OpenQASM - a quantum assembly language to
describe quantum circuits using straight-line code.
\citet{smith2016pyquil} introduced Quil - a practical quantum instruction set
architecture - which can be used to describe quantum programs with classical
control flow.
\citet{scaffcc} introduced Scaffold, an imperative programming model inspired
by classical high-level programming languages such as C and C++. It comes
with a toolchain based on the LLVM compiler infrastructure, where quantum
gates have been modeled using custom intrinsics and qubit using a custom
datatype.
\citet{QIR} introduced QIR, an LLVM based IR for the Q\# programming language.
It uses LLVM intrinsics to model quantum operations and memory semantics to
model qubit arrays.
\citet{mccaskey2021mlir} have prototyped common quantum assembly languages in
MLIR. It models qubits using memory semantics - similar to QIR.
The above IRs use memory semantics for qubits and side-effecting quantum
operations, which make reusing existing classical compiler optimizations
challenging.
\citet{xacc_2020} introduced XACC - a System-Level Software Infrastructure for
Heterogeneous Quantum-Classical Computing.
Cirq is a framework by Google Quantum AI to represent and transform
quantum circuits~\cite{cirq_developers_2021_4586899}.
\citet{aleksandrowicz2019qiskit} introduced Qiskit, which stores circuits
using an in-memory DAG representation.
The use of in-memory representations makes scaling and reuse of existing
compiler infrastructure challenging.
\citet{altenkirch2005functional} introduced QML - a functional language for
finite quantum programs QML with an accompanying compiler based on Haskell.
It uses first order strict linear logic to describe quantum computation.
\citet{Green_2013} introduced Quipper - a scalable, functional, higher-order quantum programming
language. It is geared towards a model of computation that uses a classical
computer to control a quantum device.
Silq is a high-level quantum language with safe uncomputation and intuitive
semantics~\citet{silq2020}. It has a strong static type system to enforce
physical constraints.
We believe that QSSA is the ideal optimizing IR for such languages, as SSA is
effectively a functional language~\cite{appel1998ssaisfunctional}, and it
brings a host of optimization opportunities to them. \citet{schuiki2020LLHD}
introduce an SSA-based multi-level IR for hardware description languages. It
is used to design classical circuits, which have some similarities to quantum
circuit design and analysis.
We acknowledge concurrent work by \citet{ittah2021enabling} which designs a
similar IR for optimizing quantum programs.

\section{Conclusion}

Prior work on developing IRs for hybrid quantum-classical programs
has focused on adapting classical compiler technology to the quantum
regime. However, the translations have ignored the necessary adaptations
necessary to exploit SSA's representational power.  We present QSSA, SSA-based
IR for representing hybrid quantum-classical programs. We model all quantum
operations using side-effect SSA value semantics. This allows us to reuse well
known SSA analyses and optimizations, thereby allowing us to leverage the
extensive research into SSA optimization.  We demonstrate the effectiveness of
our approach by evaluating our compiler infrastructure against the QASMBench and
IBM Quantum Challenge Datasets, where we achieve performance parity with Qiskit's
optimization pipeline. QSSA enables quantum compilation technology to use
decades of SSA research, such as tools like Alive \cite{lopes2015provably} to
verify quantum optimizations, and stochastic superoptimizers like STOKE
\cite{schkufza2013stochastic} to discover novel quantum optimizations.
In conclusion, QSSA allows us to represent, analyze, and transform quantum
programs using the robust theory of SSA representations, bringing quantum
compilation into the realm of well-understood theory and practice.

\bibliography{references}

\end{document}